\documentclass[oneside]{article}
\usepackage[T1]{fontenc}
\usepackage{authblk}
\usepackage{float}
\usepackage{amsmath,bm}
\usepackage{graphicx}
\usepackage{xfrac}
\usepackage[english]{babel}
\usepackage{lmodern}
\usepackage[T1]{fontenc}
\usepackage[latin9]{inputenc}
\usepackage{mathtools}
\usepackage{graphicx}
\usepackage{esint}
\usepackage[unicode=true,
 bookmarks=false,
breaklinks=false,pdfborder={0 0 1},backref=false,colorlinks=false]
 {hyperref}
\usepackage[a4paper, total={210mm,297mm},margin=2.0cm]{geometry}

\usepackage{datetime}
\newdate{date}{14}{07}{2018}

\title{\textbf{Mesoscale modelling of soft flowing crystals}}

\author[1]{Andrea Montessori}

\author[2]{Marco Lauricella}

\author[3,4]{Sauro Succi \thanks{Electronic address: \texttt{s.succi@iac.cnr.it}; Corresponding author}}

\affil[1]{Department of Engineering, University of Rome \textquotedbl{}Roma Tre,\textquotedbl{} Via della Vasca Navale 79, 00141 Rome, Italy}

\affil[2]{Istituto per le Applicazioni del Calcolo CNR, Via dei Taurini 19,
00185 Rome, Italy}

\affil[3]{Center for life nanoscience at la Sapienza, Istituto Italiano di Tecnologia, viale Regina Elena 295, 00161, Rome, Italy}

\affil[4]{Institute for Applied Computational Science, Harvard John A. Paulson School of Engineering And Applied Sciences, Cambridge, MA 02138, United States}

\date{\displaydate{date}}

\begin{document}

\maketitle

\begin{abstract}

We outline the main ideas behind the numerical modelling of soft-flowing crystals, with special attention
to their application to microfluidic devices for the design of novel mesoscale porous materials. 

\end{abstract}

\section{Introduction}

Recent years have witnessed a rapid surge of microfluidic experiments, showing the capability to
produce highly ordered emulsions, namely a flowing collection of liquid droplets 
(say oil in water) characterised by a highly ordered and uniform, crystal-like structure.
Whence the name of {\it Soft Flowing Crystals} (SFC).

Besides their inherent interest as a peculiar state of complex flowing matter, SFC's lend themselves to many potential 
applications, such as scaffolds for tissue engineering or highly regular porous materials for catalysis,
to mention but two major ones \cite{costantini2015microfluidic,shum2011multicompartment,hamley2009ordering}. 

This has spurred a major experimental activity in the field of microfluidics, 
many device variants being proposed with the aim of maximizing the outflow rate of  the soft crystal,  while retaining its ordered structure.

Clearly, these are conflicting constraints, since the net macroscopic motion of the crystal 
structure leads to deformations and eventually to the very rupture  of the flowing crystal.

Whence the intense search for optimal solutions, based on combinations of geometrical 
and operational parameters of the corresponding devices, T-junctions, flow-focusers, 
step-emulsifiers, being only a few of the current designs under exploration.


\maketitle

Notwithstanding such surge of experimental activity, the numerical description of SFC, 
as well as the identification of optimal operating conditions of the aforementioned devices, is still in 
its early stage. In this paper, we wish to describe the main ideas behind the mesoscale 
modelling of SFC devices and present some current applications to flow-focusers and step-emulsifiers.  

\section{Physics of SFC devices}

By soft-flowing crystal (SFC), we imply the ordered state of matter consisting of a 
collection of mesoscale droplets, distributed in space according to a regular pattern (the crystal). 
If the droplets are spheres of equal diameter, the crystal is said to be mono-disperse, otherwise it is poly-disperse.

\begin{figure}[!ht]
\begin{centering}
\includegraphics[width=0.6\columnwidth]{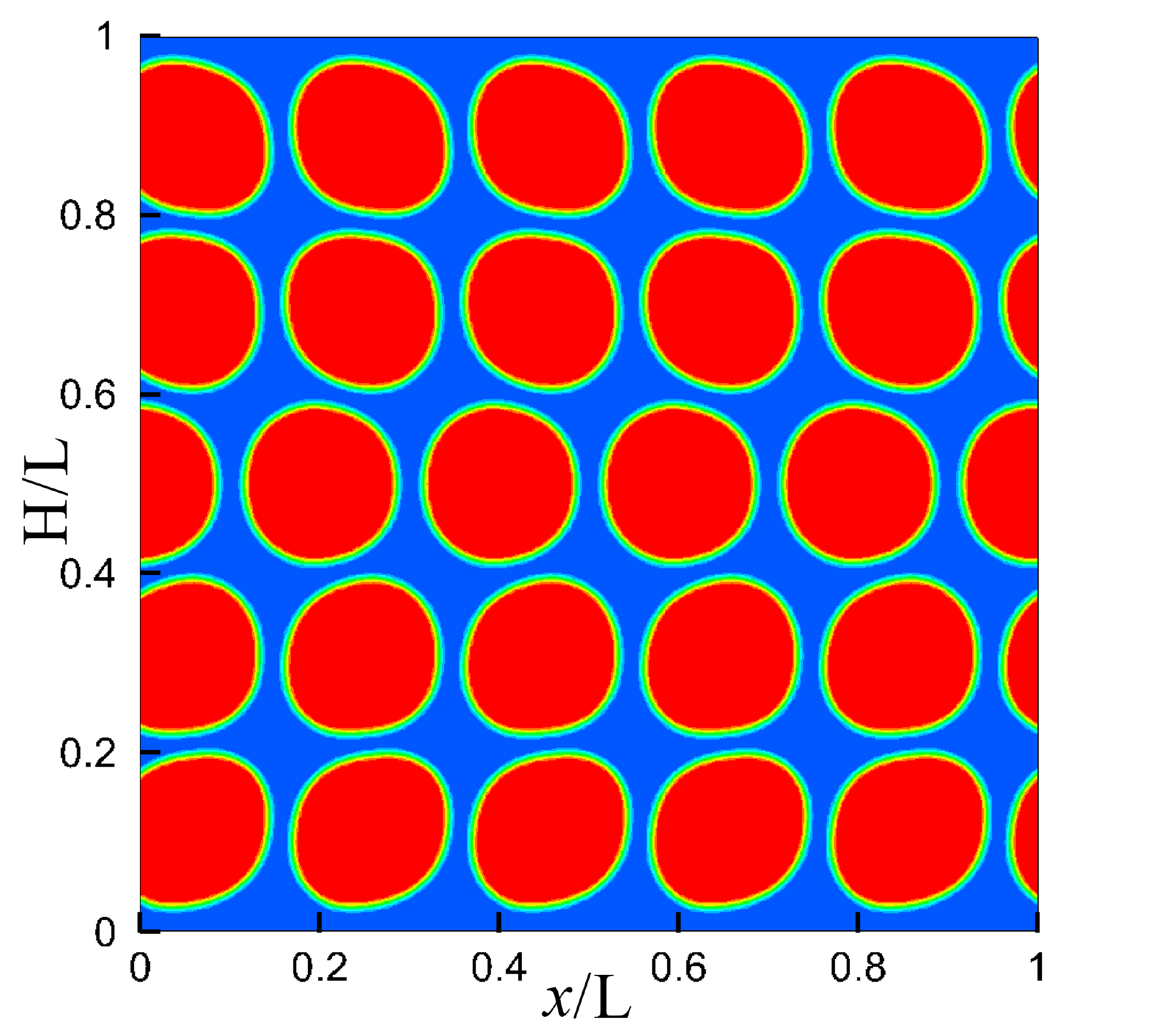}
\par\end{centering}
\caption{Snapshot of pressure-driven SFC in a planar channel at a packing fraction $\psi = 0.64$.
\label{Fig:droplet_train}
}
\end{figure}

At this stage, droplets can be regarded as upscaled version of atoms, by some five orders of magnitude, 
say from nanometers to hundreds of microns.
The crystal is soft, because droplets can deform under the effect of shear and
mutual collisions at high packing fractions (ratio of droplets volume to the total volume).
The crystal is also flowing because proves capable supporting net coherent motion of 
the droplets, as a single macroscopic entity (see Fig. \ref{Fig:droplet_train}).

Why are SFC an interesting of state of matter?

From a fundamental point of view, they offer an intriguing example of a mesoscale 
hybrid between solid and liquid state, whose rheological properties are key to many practical 
applications in material science, engineering, biology and medicine,
from tissue engineering, to catalysis and many others \cite{hu2014directed,chremos2010ultra,hamley2009ordering,cheng2004nanostructure}.  

It turns out that such mesoscale porous materials can be manufactured quite reliably using a 
variety microfluidic devices and by regulating a few macroscopic operational parameters, 
such as the geometrical set up and the
flow rates of the various components \cite{garstecki2006flowing}. 
For the sake of concreteness, we shall refer to a binary mixture of oil droplets (the dispersed phase)
in water (the continuum phase).

The mandate of the micro device is to produce substantial amounts of SFC's with a high degree of regularity 
and monodispersity, and droplet diameters in the order of 10 to 100 microns, depending on the 
specific application.

Next, some notation.
The volume fraction (or packing fraction) is defined as
\begin{equation} 
\psi = \frac{V_d}{V}   
\end{equation}
where $V_d= N \pi D^3/6$ is the volume (area, in two space dimensions) occupied 
by $N$ droplets of diameter $D$ and $V=H L W$ is the total volume of the device, 
a planar channel of height $H$, length $L$ and width $W$.

The inter-droplet separation is $d=n^{-1/3}$, $n=N/V$ being the number density of the droplets.

The inter droplet gap $g=d-D$ defines the interface separation and the condition $g=0$ denotes 
maximum-packing, i.e. droplets come into contact.
The outflow rate (number of droplets per unit time) of the dispersed phase 
is $Q_d = n U H W$, where $U$ is the outflow velocity.

The goal is to maximize $Q_d$ with the desired value of the droplet diameters.

The droplet diameter is a function of the geometrical parameters, typically, in a flow-focuser, the crossflow 
height of the channel, $H$, and the ratio of the inflow rates of the dispersed 
to the continuum phase $q= Q_d^{in}/Q_c^{in}$. 

In addition, it also depends on a series of physico-chemical parameters,
such as the viscosity ratio $M = \mu_d/\mu_c$, the surface tension $\sigma $, which 
we denote generically as $p$.

In fully generality:
\begin{equation}
D/H = f(q, p)
\end{equation} 

Hence, the optimization problem is to find the set of parameters, $p$, such that $Q_d$ is
maximum at the desired value of $D$.

Given the mesoscale nature of SFC's, mesoscale methods based on kinetic theory appear 
well suited to their numerical modelling, hence in the sequel we shall outline the main features of a 
particularly popular mesoscale method, known as Lattice Boltzmann (LB).

\section{Lattice Boltzmann Models of SFC's}

The lattice Boltzmann model is a mesoscale techniques based on the evolution of a discrete Boltzmann distribution on a uniform lattice with sufficient symmetries to recover macroscopic fluid dynamics in the large-scale limit, i.e. at scales
much larger than the molecular mean free path.
The method has made the object of a vaste literature, including a number of exhaustive monographs \cite{succi2018lattice,kruger2017lattice}, hence in this
work we shall only lay down the basic equation, leaving all the details to the original literature.

The LB reads as follows:

\begin{equation} \label{LBE}
f_{i} \left(\vec{x}+\vec{c}_{i}\Delta t,\,t+\Delta t\right) =f_{i}\left(\vec{x},\,t\right)+\Omega_{i}( f_{i}\left(\vec{x},\,t\right)) +\vec{F}_i\Delta t
\end{equation}
where $f_{i}$ is the discrete distribution function, representing
the probability of finding a fluid particle at position $\vec{x}$ and time
$t$ with discrete velocity $\vec{c}_{i}$ . 
The lattice time step is taken equal to 1, and $i$ the index spans the lattice discrete 
directions $i = 0,...,b$, where $b=8$ for a standard 
two dimensional nine speed lattice (D2Q9) while is $18$ or $26$ for three-dimensional lattices (D3Q19 and D3Q27). 

The left-hand-side represents the free molecular streaming, while the right hand side codes 
for the collisional interactions and the effect of external/internal forces acting on the fluid molecules.

The fluid density $\rho$ is given by the zeroth order moment of the distribution functions
\begin{equation}
\rho\left(\vec{x},\,t\right) = \sum_i f_{i}\left(\vec{x},\,t\right),
\end{equation}
while the linear momentum $\rho \vec{u} $ is defined by the first order moment of the set of distributions:
\begin{equation}
\rho \vec{u} = \sum_i f_{i}\left(\vec{x},\,t\right) \vec{c}_{i}.
\end{equation}

In order to model SFC's, LB schemes must support the following features:

i) A small surface tension, to facilitate the onset of complex interfaces with high surface/volume ratios;

ii) Positive disjoining pressure, to delay (suppress) coalescence and sustain long-lived states with high S/V ratios;

iii) Non-ideal equation of state, whenever dynamic phase transitions must be included.

These physical requirements must be confronted with two well-known lattice artefacts, spurious currents and
thick interfaces. The former refers to the spurious recirculation patterns which set up around the droplet due 
to lack of symmetry of high-order tensors describing the capillary forces/ Such spurious patterns 
grow in size and intensity with the density contrast between the two phases, as well as with the 
speed of the droplet.

Finite-width interfaces, typically a few lattice spacings, are an inevitable shortcoming
of any grid-based (Eulerian) treatment of the interface, as opposed to interface-tracking Lagrangian techniques.

Both artefacts need to be minimised when designing a suitable LB for SFC.

To place this statement in a concrete perspective, it is worth taking a close look at the typical
length-scales of the problem and the ensuing computational requirements.
 
\subsection{Computing across SFC scales}

For the sake of concreteness, we refer to a microfluidic channel of 
length $L$, height $H$ and width $W$, ordered as follows $H \ll W \ll L$.

The channel hosts a collection of droplets (dispersed phase, say oil) co-flowing 
with a solvent (continuum phase, say water), of diameter $D$, whose centres are
spaced a mean distance $d$.
As previously defined, the gap between two droplets, $g$, is given by $g=d-D$, and sets
the relevant scale for near-contact interactions (NCI).

Finally, each droplet is separated from the continuous phase by an interface of width $w$.

The typical SFC ordering relevant to microfluidic experiments is:
\begin{equation}
w \ll h \ll D \sim d \ll H \ll W < L
\end{equation}
We shall further assume high packing fractions and droplet diameters comparable with the 
channel height $H$, so that only a few rows of droplets, say up to five, are free 
to flow along the micro channel.

For many applications a typical value is $D \sim 100$ microns, so that we take 
$H \sim 500$ microns, corresponding to at most five rows of flowing droplets within the micro-channel.
For typical volume fractions of SFC's, droplets are at near contact, say $g < 10$ nm, while the
interface width is of the order of $w=1$ nm or less, basically the range of molecular interactions, $r_0$.

With these numbers in mind, a typical LB simulation working at a spatial resolution $\Delta = 1$ micron, requires
a grid of the order of $N_G = 500 \times   5000 \times 5000$, namely  $12.5$ billion grid points.
This is close to the largest scale present-day LB simulations.

The potential problem, though, is that with the above set up, the NCI's are largely under-resolved, 
i.e. $\Delta/g = 100$ , while a fully-resolved simulation, by definition, commands $\Delta/g < 1$.    

Hence, a fully resolved simulation would require about $100^3$ more grid points, and $100^4$
more compute power, which is far beyond the capabilities of any foreseeable computer, a point to
which we shall return in the closing section of this paper.

The under-resolution issue is potentially serious because near-contact interactions, 
disjoining  pressure, lubrication forces and the whole family of dispersion forces, may develop 
singularities, i.e. divergences in the "UV" limit   $g/D \to 0$.

Clearly, such divergence is an artefact of the continuum representation and is naturally regulated by the finite range of the interactions, say $r_0$.

However, we note that, besides being computationally unviable, $r_0$ is not the relevant
regularization scale, because deformability effects kick in at much
larger scales than the atomistic one.
As a result, the question on the reliability of the numerical regularization due to the finite lattice
spacing, versus the physical one due to deformability effects, stands out
as one of most pressing problems in the numerical modelling of SFC's.

Failing a fully-fledged coarse-graining theory, two options arise:

1) Hard-way: Increase the grid resolution; 
2) Soft-way: Develop new coarse-grained LB models incorporating NCI effects.

Route 1) consists of enhancing the resolution selectively, 
i.e. in the gap region between two droplets, while leaving
an ordinary spacing elsewhere (inside the droplets). 
This can be achieved in many ways, by local-grid refinement, 
adaptive unstructured meshes, as well
as Lagrangian methods. 
Improvement is guaranteed, but comes at a high cost in programming complexity, due to the 
non-trivial (dynamic) data structures involved in the actual implementation, 
especially in combination with parallel computing.

Route 2) is much simpler on the programming side, but always under question as to its physical reliability.

In the following we shall provide a brief account of route 2).

\section{Coarse-graining near-contact interactions}

Since SFC's operate at close packing, near-contact interactions (NCI) play a major role on their rheology.
In particular, the question is to what degree their details affects the regularity 
and uniformity of the flowing crystal.
Such NCI embrace a vaste collection of dispersion interactions, lubrication, van der Waals, steric, possibly  even Casimir-like forces. 
Most of these forces display short-range singularities which are, in principle, a source of concern for the LB representation.

However, for deformable particles, such singularities is allegedly regulated at the scale defined by the
typical deformation length of the droplet, namely
$$
\delta \sim D \Sigma/E
$$
where $\Sigma$ is the normal stress and $E$  is the Young modulus of the droplet.

In fact, most formal coarse-graining procedures would turn the 
singular lubrication force $F(g) \sim 1/g$ into
a smoothed version $F_{\delta}(g)$, recovering the $1/g$ decay at 
large separations $g /\delta \gg 1$ and turning into a constant of 
order $1/\delta$ in the short-range limit $g/\delta \ll 1$.

It is important to note that, regardless of the details of the coarse-graining procedure, for
droplet diameters in the order of $D \sim 100$ microns, a mere $10$ percent relative deformation implies
$\delta \sim 10$ micron, hence well within the mesoscale regime suitable for Lattice Boltzmann modelling.

Therefore, a necessary condition for LB to correct capture deformability effects 
is to resolve the deformation scale, i.e. $\Delta/\delta \ll 1$.   
Next, we recall that LB interfaces come with a finite width, $w_{\text{LB}}$ of a few lattice spacings,
typically $w_{\text{LB}} \sim (3 \div 5) \Delta$ (not to confuse $w_{\text{LB}}$ with 
the physical width $w$, which is obviously much smaller).

Hence, the above condition turns into a more stringent one, namely
\begin{equation}
\label{DEF}
Cn \equiv \frac{w_{\text{LB}}}{\delta} \ll 1
\end{equation}
where $Cn$ as the Cahn number of the droplet.

\begin{figure}[ht]
\begin{centering}
\includegraphics[width=0.9\columnwidth]{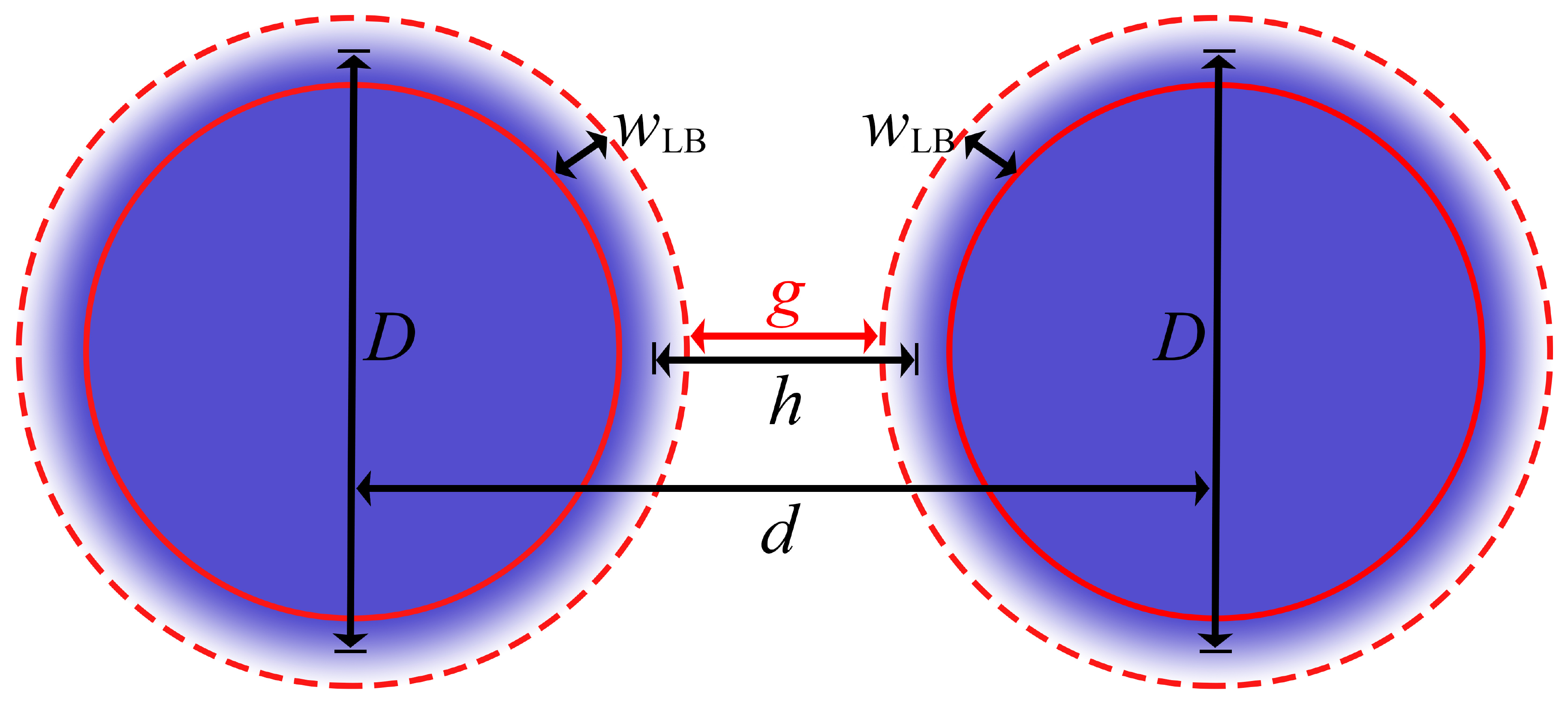}
\par\end{centering}
\caption{\label{sketchdrops}Sketch showing two LB droplets of diameter $D$ placed at mutual distance $d$, 
with diffuse interface of finite width equal to the lattice spacing $w_{\text{LB}}= (3 \div 5) \Delta$.
The length $h$ is defined as the distance between the midpoints
of the diffused interfaces, while the gap $g$ denotes the distance between the extreme
points of the two interfaces. Hence $g=h-w_{\text{LB}}$, so that $g=0$ denotes contact and $g<0$ coalescence.
\label{Fig:f_disp}
}
\end{figure}

It is of some interest to show that formal averaging over a finite width leads 
to a regularisation at the length scale of the interface width. 
To this purpose, let us assume that the force between the finite-width interfaces
can be expressed as a linear summation upon all widths $r_0 \le w' \le w_{\text{LB}}$.  
Taking a singular form $F(h;w')=1/(h-w')$ and defined $g=h-w_{\text{LB}}$, the integral over the range of widths 
from $r_0$ up to $w_{\text{LB}}$ delivers (see Fig. \ref{sketchdrops}):

\begin{equation}
\bar F(h;w_{\text{LB}}) = \frac{1}{w_{\text{LB}}-r_0}  \int_{r_0}^{w_{\text{LB}}} F(h,w') dw' =
\frac{1}{w_{\text{LB}}-r_0} \log (\frac{g+ w_{\text{LB}}}{g+r_0})   
\end{equation}

In the far-field limit $w_{\text{LB}}/g \to 0$, this returns $1/g$, i.e the bare singularity
between rigid particles, as it should be.
In the opposite, near-field, limit $g/w_{\text{LB}}\to 0$, we obtain
$\bar F(h;w_{\text{LB}}) \sim 1/(w_{\text{LB}}-r_0) \; log (w_{\text{LB}}/r_0) \sim O(1/w_{\text{LB}})$, the expected regularisation effect.

\subsection{Soft droplets: the effects of deformability}

As mentioned in the introduction, droplets are susceptible of 
appreciable deformation under shear, the
dimensionless parameter in control of their deformability being the Weber number, 
defined as:
\begin{equation}
We = \frac{\rho U^2D}{\sigma}
\end{equation}
Deformability effects kick in at $We > 1$ and are 
expected to offer a natural regularisation of NCI's
at scales well above the atomistic one.

Typical operational parameters are 
(in MKS units) $U=10^{-2}$, $D=10^{-4}$, $\sigma=0.07$ and $\rho=10^3$, which
deliver $We = 10^{-4}$.  Hence, interaction with the 
continuum phase does not lead to appreciable deformations.

This is however not the case at the injection nozzle, 
where the injected droplets literally bump into 
their forerunners, converting kinetic into elastic deformation energy. 
This mechanism is in operation whenever droplets collide with each other.

Here, the relevant control parameter is the Bond number, defined as
$$
Bo = \frac{\Delta \rho a D^2}{\sigma}
$$ 
where $a$ is the acceleration due to any generic volume force acting upon the droplet.

Rigid particles associate with the limit $Bo \to 0$, i.e. 
nominally infinite surface tension.

With typical experimental parameters (oil-water system, MKS units), 
$\Delta \rho \sim 100$, $D=10^{-4}$ and $\sigma \sim 0.004$ , 
$a \sim 10^3$ (injection at 1 m/s every millisecond), 
we obtain $Bo \sim 0.25$, indicating that deformability
effects cannot be ignored at the injection inlet. 

Indeed, both experiments and simulations show clear evidence 
of substantial droplet deformations of the order 
of $\delta /D \sim 0.1$, which imply elastic displacements $\delta D$ of the order
of microns and above, hence well above the atomistic scale and within reach of a mesoscale treatment.  

This is of course a purely heuristic argument, but one which matches experimental 
and numerical (see later) observations.

As a result, failing a rigorous many-body theory of soft particles, 
we proceed by enriching 
LB models with features optimizing its ability to reproduce experimental results, 
what we call a rule-driven approach.

Many LB variants can be devised to this purpose, 
but in the sequel we provide a few details on the one
which, so far, best served the above purpose, namely the regularised, 
multi-range, colour-gradient LB.  

\subsection{Multi-range Regularized Colour-Gradient \label{color}} 

The colour gradient LB for multicomponent flows is briefly described below:

Two sets of distribution functions are needed to track the evolution of the two fluid components, which occurs via a streaming-collision algorithm:

\begin{equation} \label{CGLBE}
f_{i}^{k} \left(\vec{x}+\vec{c}_{i}\Delta t,\,t+\Delta t\right) =f_{i}^{k}\left(\vec{x},\,t\right)+\Omega_{i}^{k}( f_{i}^{k}\left(\vec{x},\,t\right)),
\end{equation}

the index $k$ spanning over the components of the fluid system.

In these models, the collision operator is usually split into three parts \cite{gunstensen1991lattice,leclaire2012numerical}: 

\begin{equation}
\Omega_{i}^{k} = \left(\Omega_{i}^{k}\right)^{(3)}\left[\left(\Omega_{i}^{k}\right)^{(1)}+\left(\Omega_{i}^{k}\right)^{(2)}\right].
\end{equation}

In the above, $\left(\Omega_{i}^{k}\right)^{(1)}$ stands for the standard collisional relaxation, $\left(\Omega_{i}^{k}\right)^{(2)}$ is the so-called perturbation step, which 
contributes to the build up of an interfacial tension. Finally, $\left(\Omega_{i}^{k}\right)^{(3)}$ is the recoloring step, which mimics the segregation 
between species, so as to minimize their mutual diffusion.

The perturbation operator can be constructed directly by exploiting the concept of the continuum surface force \cite{brackbill1992continuum}, so as to reproduce 
the correct form of the capillary-stress tensor \cite{landau1959course}.

Firstly, the perturbation operator must satisfy the following  conservation constraints:

\begin{eqnarray} \label{consconstr}
\sum_i \left(\Omega_{i}^{k}\right)^{(2)}=0 \\
\sum_i \left(\Omega_{i}^{k}\right)^{(2)} \vec{c}_i=0
\end{eqnarray}

By performing a Chapman-Enskog expansion, it is possible to show that the hydrodynamic limit of Eq.\ref{CGLBE} is 
a set of equations for the conservation of mass and linear momentum:

\begin{eqnarray} \label{NSE}
\frac{\partial \rho}{\partial t} + \nabla \cdot {\rho \vec{u}}=0 \\
\frac{\partial \rho \vec{u}}{\partial t} + \nabla \cdot {\rho \vec{u}\vec{u}}=-\nabla p + \nabla \cdot [\rho \nu (\nabla \vec{u} + \nabla \vec{u}^T)] + \nabla \cdot \bm{\Sigma}
\end{eqnarray}

where $p=\sum_k p_k$ is the pressure and $\nu=c_s^2(\tau-1/2)$ is the kinematic viscosity of the mixture.

The divergence term in the momentum equation ($\nabla \cdot \bm{\Sigma}$) is given by:

\begin{equation}
\bm{\Sigma}=-\tau\sum_i \sum_k\left(\Omega_{i}^{k}\right)^{(2)} \vec{c}_i \vec{c_i}
\end{equation}

The surface stress boundary condition at the interface between two fluids 
can be expressed as \cite{landau1959course,brackbill1992continuum}:

\begin{equation}
\mathbf{T}^1 \cdot \vec{n} - \mathbf{T}^2 \cdot \vec{n} = \sigma (\nabla \cdot {\vec{n}}) \vec{n} - \nabla \sigma
\end{equation}

where, $\mathbf{I}$ is the identity tensor, $\sigma$ is the surface tension coefficient, $\vec{n}$ is the unit normal to the interface, $\mathbf{T}=-p\mathbf{I} + \rho\nu(\nabla\vec{u} + \nabla \vec{u}^T)$ is the stress tensor of the $k-th$ component and $\nabla \cdot \vec{n}$ is the local curvature of the fluid interface.

Thus, the local stress jump at the interface can be induced by adding a volume interfacial force $\vec{F}(\mathbf{x},t)$ \cite{liu2012three}:

\begin{equation}\label{force}
\vec{F}(\vec{x},t)= \nabla \sigma \delta_{I} - \sigma (\nabla \cdot \vec{n}) \vec{n} \delta_{I}
\end{equation}

In the above, $\delta_{I}=\frac{1}{2} | \nabla \rho^N | $ is an index function localizing the force explicitly 
on the interface and $\rho^N=\frac{\rho^1 - \rho^2}{\rho^1 + \rho^2}$ is the phase field \cite{liu2012three}.
The normal to the interface can be approximated by the gradient of the phase field, $\mathbf{n}= \nabla \rho^N/|\nabla \rho^N|$.

Since the perturbation operator is responsible for generating interfacial tension, the following 
relation must hold:

\begin{equation} \label{SeqF}
\nabla \cdot \bm{\Sigma}= \vec{F}
\end{equation}

By choosing \cite{reis2007lattice} $\left(\Omega_{i}^{k}\right)^{(2)}= A_k |\nabla \rho^N|\left[w_i \frac{(\vec{c}_i \cdot \nabla \rho^N)^2}{|\nabla \rho^N|^2} -B_i \right]$ and substituting into \ref{consconstr} and \ref{SeqF} and by imposing that the set $B_i$ must satisfy the following isotropy constraints:

\begin{eqnarray}
\sum_i B_i= \frac{1}{3} \; \sum_i B_i \vec{c}_i=0 \; \sum_i B_i \vec{c}_i \vec{c}_i= \frac{1}{3} \mathbf{I}
\end{eqnarray}

we obtain an equation for the surface tension of the model:

\begin{eqnarray}
\sigma=\frac{2}{9}(A_1+A_2)\tau
\end{eqnarray}

This shows a direct link between the surface tension and the parameters $A_1$ and $A_2$.

As we pointed out, the perturbation operator generates an interfacial
tension in compliance with the capillary-stress tensor of the Navier Stokes equations 
for a multicomponent fluid system.

The perturbation operator alone, does not guarantee the immiscibility of different fluid components.
For this reason, a further step is needed (i.e. the recoloring step) to minimize the mutual diffusion between components.

Following the work of Latva-Kokko and
Rothman \cite{latva2005diffusion}, the recoloring operator for the two sets of distributions takes the following form:

\begin{eqnarray}
f_i^1=\frac{\rho^1}{\rho} f_i^* + \beta \frac{\rho^1\rho^2}{\rho^2} \cos{\phi_i} f_i(\rho,\vec{u}=0)^{eq} \\
f_i^2=\frac{\rho^2}{\rho} f_i^* - \beta \frac{\rho^1\rho^2}{\rho^2} \cos{\phi_i} f_i(\rho,\vec{u}=0)^{eq}
\end{eqnarray}

where $f_i^*=\sum_k f_i^{k,*}$ denotes the set of post-perturbation distributions, $\rho=\rho^1 + \rho^2$, $\cos{\phi_i}$ is the angle between the phase field gradient and the $i-th$ lattice vector and $f_i(\rho,\vec{u}=0)^{eq}=\sum_k f_i^k(\rho,\vec{u}=0)^{eq}$ is the total zero-velocity equilibrium distribution function.

Note that $\beta$ is the only free parameter in this equation and can be used 
to tune the interface width, thus playing the role of an inverse diffusion length
scale \cite{latva2005diffusion}.

In other words, the recolouring step aims at redistributing the colour in such
a way that the difference between the densities of the two components (i.e. $\rho^1 - \rho^2$) along the $i-th$
direction depends continuously on the angle $\phi_i$ \cite{d1995two}. In addition to the
conservation of the total density ($\rho^1 + \rho^2$) site by site,  we
also demand that $f_i^* = f_i^{1,*} + f_i^{2,*}$ . 

For applications where there is a specific need to prevent coalescence between droplets of the same 
fluids, like in emulsions and foams, a further step is required, which consists of assigning a 
new index (flavour) to newborn droplets and 
impose a "flavour-dynamic" extra-repulsion, between different flavours on top
of the mechanical repulsion due to the colour gradient and the recolouring stage.

In a 2-D simulation, six flavours are generally sufficient to prevent neighbour droplets with the same colour \cite{montessori2018regularized,dupin2006simulation}.

The LB scheme is further regularized,  by filtering out the high-order non-hydrodynamic (ghost) modes after the streaming 
step \cite{montessori2015lattice,zhang2006efficient}.
Indeed, it was noted that sizeable non-isotropic effects 
arise in the model \cite{montessori2018regularized}, 
whenever the LB scheme is under relaxed.
As a consequence, we exploit the regularization procedure
in order to recover the loss of isotropy by
suppressing the non-hydrodynamic modes.

To date, it was found that the two-belt (gradient includes second Brillouin cell), regularized, 
colour-gradient model has provided the best performance for the set of microfluidics experiments 
described in  the sequel \cite{montessori2018regularized}.

It is a blend of rule-driven and physically-driven model, where the rule-driven part
incorporates the knowledge gained by comparison with experimental results. 
For the future, it would be interesting to incorporate such knowledge in a more systematic
way, possibly via machine-learning algorithms, i.e. by presenting the algorithm with a series
of experimental flows and minimise the cost function through appropriate variation of the
parameters of the model.

\section{Applications}

Several droplet-based microfluidic chips include one or more droplet generation units in different geometrical configuration, alongside with droplet splitting/merging units.
Nowadays, such devices are common tools for producing SFC. Among these we mention microfluidic chips as
flow focusing, co-flow, T-, X-, and Y-junctions.
Here, we report few applications of the mentioned models for SFC in the mentioned microfluidic chips, 
namely the simulation of droplet formation in a standard double T- junction, in a flow-focussing device 
and in the recently proposed parallel step emulsifier device \cite{stolovicki2018throughput}.

\subsection{Double T- junction}

The double T- junction is a compact hydrodynamic geometry which 
consist of four
micro-channels with same width intersecting at right angles 
in a cross channel
configuration. The resulting microfluidic chip contains a 
droplet generation unit with an high control on the degree 
of monodispersity among droplets.
In particular, the volumetric flow rates in the
inlet channels can be easily tuned in order to drive the final 
pattern of the SFC.

\begin{figure}[ht]
\centering
\includegraphics[width=0.9\columnwidth]{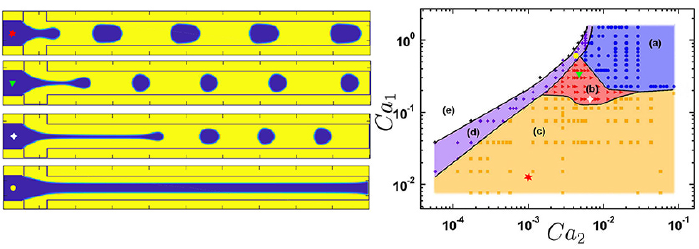}
\caption{\label{regimes}  capillary number-based flow map with flow regimes reproduced by the 
regularized colour gradient model (from the upper left to the lower left panel): Dripping, jetting 
(second and third) and tubing.
The regularized model is capable of accurately predicting the different flow regimes in a microfluidic 
flow focusing device, thus matching the viscosity ratio of the liquids employed in the 
experiments reported in \cite{cubaud2008capillary} (right panel).}
\end{figure}

We performed simulations of two component fluids modelled by the chromodynamic model discussed 
in Subsection 4.\ref{color}.
In particular, we consider a blue component (dispersed phase) and a second yellow components 
(continuous phase) with same density.
The viscosities of the two components are $\nu_c=0.167$ for the 
continuous phase and $\nu_d \sim 0.5$ for the dispersed phase. 
The viscosity ratio of the liquids are taken in order to match 
the experimental data reported in Ref. \cite{cubaud2008capillary}.
It is worth noting that the relaxation times of the blue and 
yellow components were consequently equal to
$\tau_{c}=1$ and $\tau_{d} =2$, respectively.
In particular, the larger $\tau_{c}$ value of the yellow component 
provides an under-relaxation of the LBGK models
resulting in an out of equilibrium regime which is cured by
the regularization procedure, as already discussed 
in Subsection 4.\ref{color}.

We run several simulations of the device at different Capillary number values, $Ca$,
by tuning the surface tension and the inlet velocities.
In Fig. \ref{regimes}  we report the Capillary number-based flow map with flow 
regimes observed by Cubaud et al. \cite{cubaud2008capillary}.

\begin{figure}[ht]
\centering
\includegraphics[width=0.9\columnwidth]{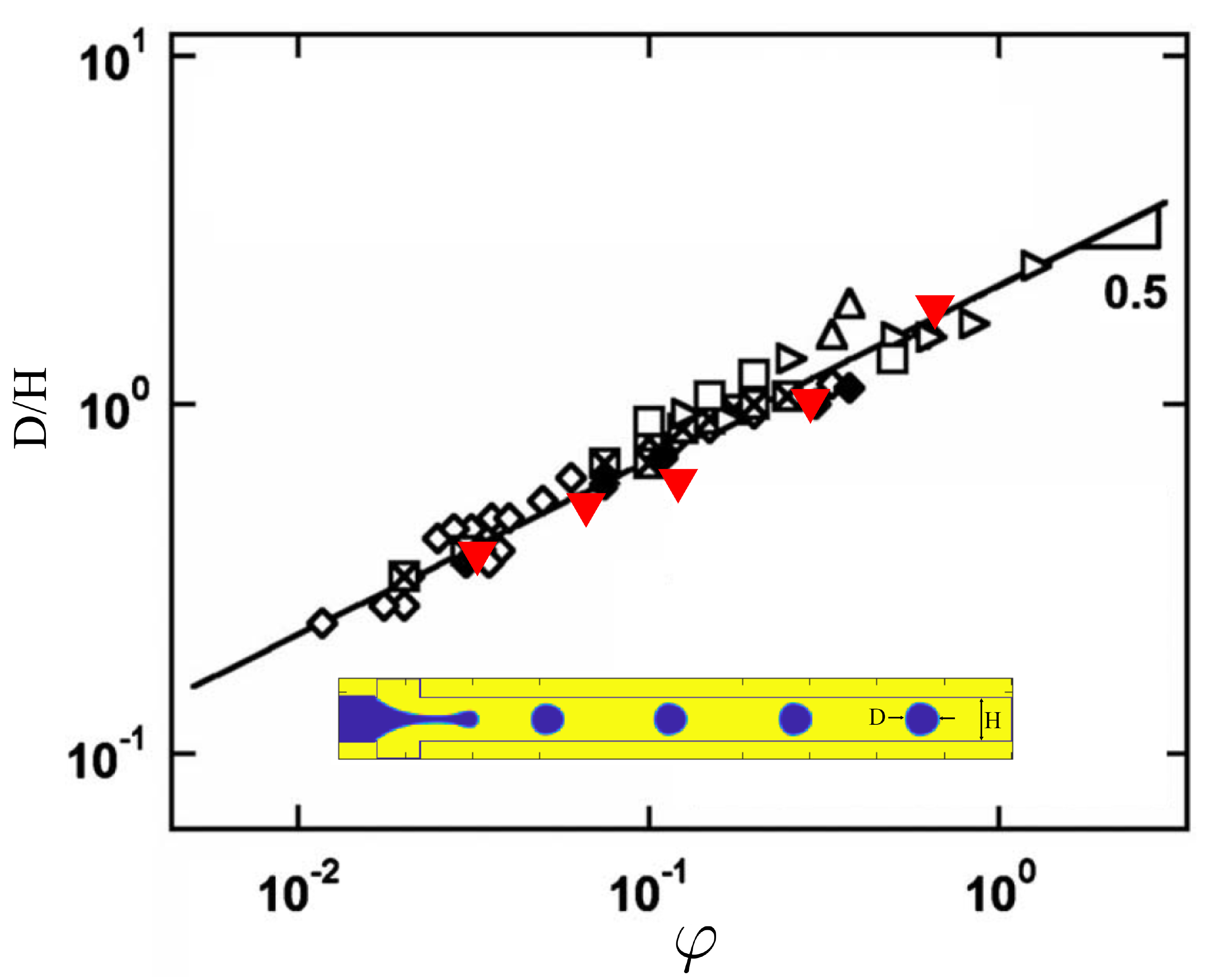}
\caption{\label{dovh} Normalized droplet diameter $ D/H$ vs the flow rate ratio $\varphi=Q_B/Q_Y$. Numerical results (Red triangles) are superimposed to the experimental curve of Cubaud et al. \cite{cubaud2008capillary}.
The diameters collapse on a single master curve, which scales with the flow rate like $Q_B/(2 Q_Y)^{0.5}$, as reported in \cite{cubaud2008capillary}.}
\end{figure}

The different flow regimes are well reproduced, showing that the actual 
the regularized colour gradient is clearly able to predict dripping, 
jetting and tubing flow configurations
at different Capillary numbers in such
microfluidic flow-focusing device.
Next, denoted $H$ the height of the microfluidic channel, we measure
the normalized droplet diameter $D/H$ versus the flow rate ratio ($\varphi=Q_d/Q_c$).
Hence, we compare our results with experimental data,
finding a good agreement with Ref. \cite{cubaud2008capillary}.
In particular, we observe a single master curve 
containing both the numerical and experimental points, which
scales with the flow rate like $Q_d/(2 Q_c)^{0.5}$ (see Fig. \ref{dovh}).

\subsection{Flow focusing}

Nowadays, a typical microfluidic configuration for the SFC 
production is the flow focusing.
The main difference from the previous double T- junction is the replacement 
of one micro-channel with an orifice, which restricts the liquid flow into the outlet channel.
As a consequence, the hydrostatic pressure increases upstream of and 
in the orifice, providing a 
squeezing effect of the immiscible thread \cite{garstecki2005mechanism}.

\begin{figure}[ht]
\centering
\includegraphics[width=0.9\columnwidth]{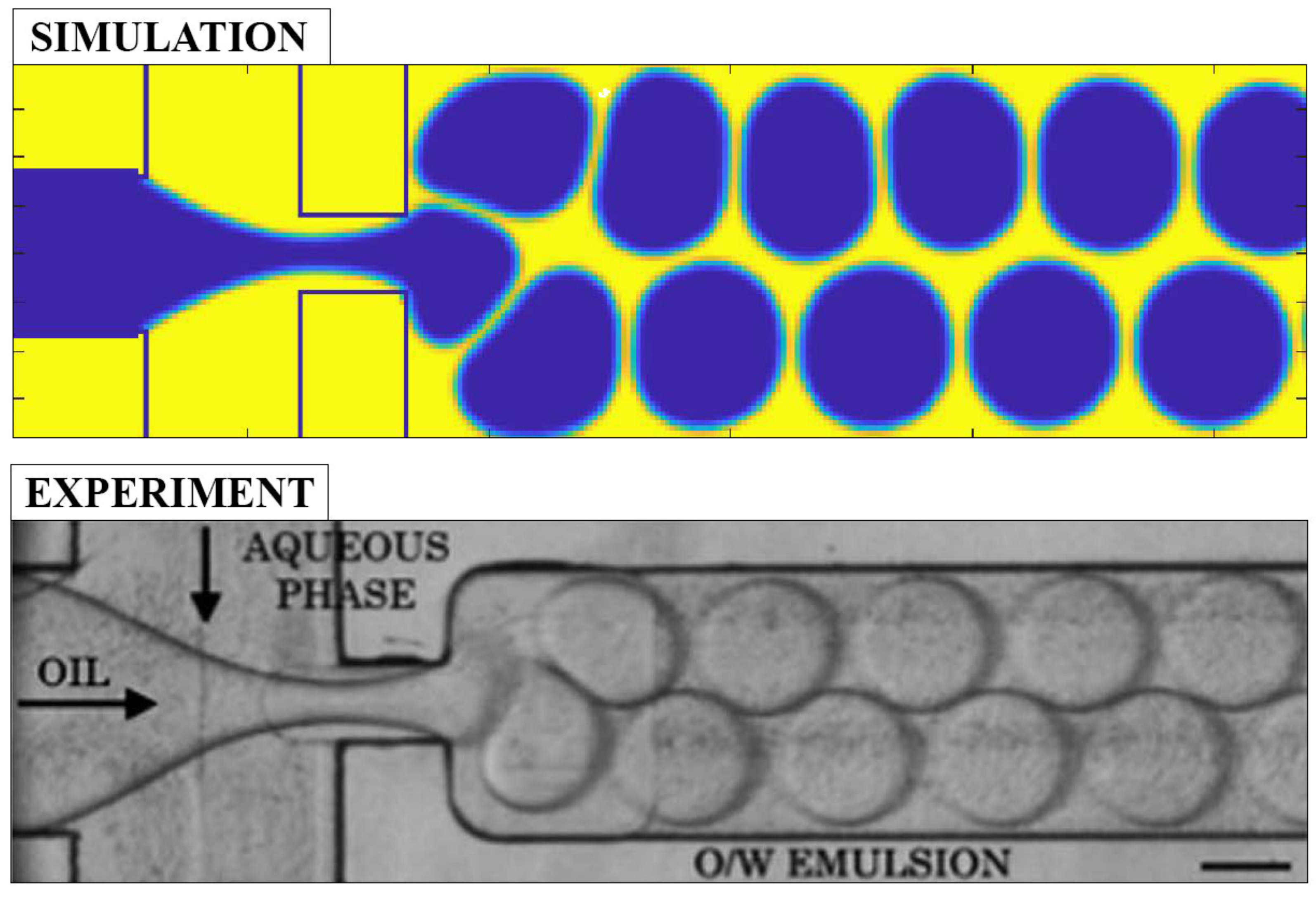}
\caption{\label{owemulsion} Prospective application of regularized colour gradient model augmented with arrested coalescence algorithm.
The model allows for stable simulation of mono-disperse droplets. This opens the way to the simulation of mono-dispersed oil-water emulsions. Future applications of this model will allow to  identify optimal operational regimes, capable of delivering droplet configurations of high regularity, both in size and connectivity. Lower panel containing the experimental data as reported from Ref. \cite{costantini2014highly} }
\end{figure}

The color-gradient model was used to simulate an oil/water emulsion 
in a flow-focusing device.
In order to obtain a mono-dispersed emulsion, 
we add an algorithm aimed at suppressing coalescence to the regularized colour 
gradient approach.
The algorithm promotes the separation between the droplets of the dispersed phase (oil),
in similarity to that one reported in \cite{dupin2006simulation}.
In particular, each newly formed droplet changes 
its colour as soon as the break up occurs.
Thus, the new droplet (blue component 
in fig \ref{owemulsion}) becomes immiscible with both the ambient 
fluid (yellow component) and its neighbour droplets in the channel.
In order to detect the drop break we monitor the dispersed phase density profile of the jet
along the horizontal central axis of the domain. 
The appearance of a zero in the density profile signals that the droplet 
separation has occurred. 
Hence, a new colour is assigned to the newly formed droplet, 
keeping the same physical properties of the dispersed phase 
(viscosity and surface tension). 
Note that the number of different colours is finite and it 
depends on the system under investigation.
In this case only five different colours are sufficient
to prevent coalescence between neighbouring droplets.
Thus, computing time and memory are virtually independent of the number 
of simulated droplets \cite{dupin2006simulation}.
It is also important to highlight that the surface tension 
between the dispersed droplet is tuned in order to guarantee 
a contact angle of $180^\circ$ between the droplets, ensuring
a complete suppression of the coalescence. 
As shown in Fig. \ref{owemulsion}, the model allows for stable simulation of mono-disperse 
droplets in good agreement with the experimental data reported in Ref. \cite{costantini2014highly},
providing an efficient tool to simulate mono-dispersed oil-water emulsions. 
This opens up new chances to investigate droplet configurations 
with high regularity, both in size and connectivity, which are typical in SFC,  
identifying the optimal operational regimes for their production. 

\subsection{Parallel step emulsifiers}

A new class of parallel step emulsification 
devices, called volcano \cite{stolovicki2018throughput}, has been recently proposed
as an efficient tool to produce emulsions at high generation rate. 
The main idea of such devices is to exploit the buoyancy effects for 
preventing the obstruction of the nozzles from the droplets.
Further, these devices are expected to increase remarkably the 
mono-dispersity of water/oil or oil/water emulsion,
which is highly useful for several manufacturing process.

\begin{figure}[ht]
\centering
\includegraphics[width=0.96\columnwidth]{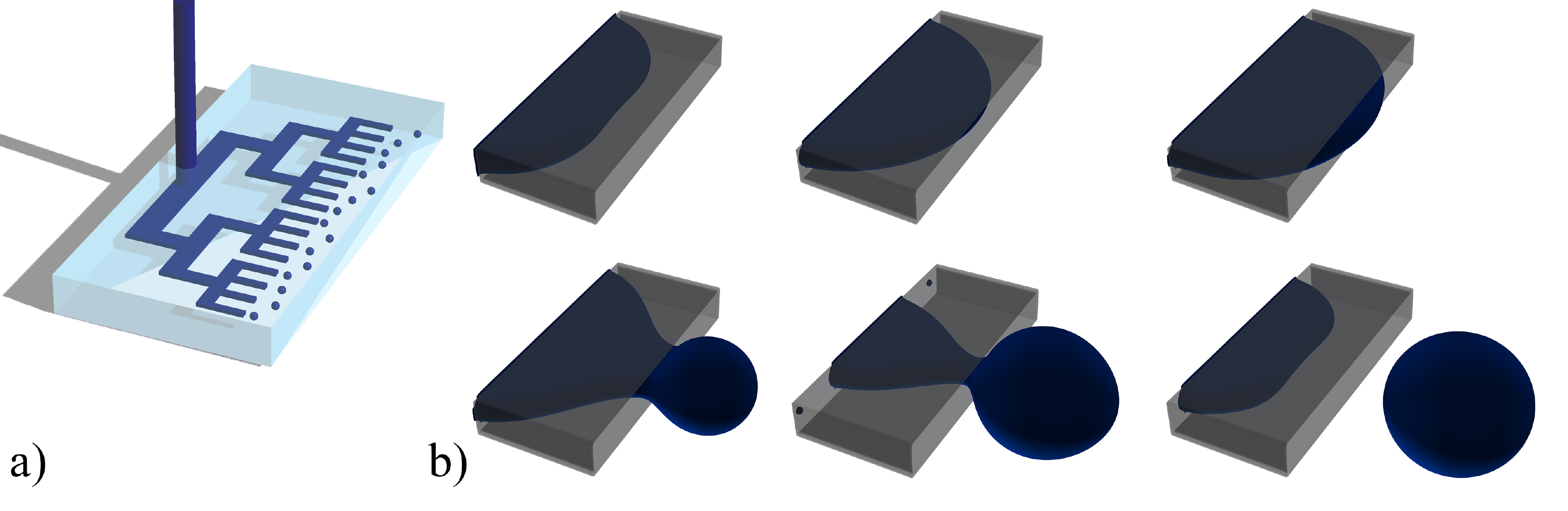}
\caption{\label{sequencevolc}  a) Sketch of the vulcano device. b) droplet breakup sequence. 
The width of the channel $W=700 \mu m $ and the  
height $H= 70 \mu m$ ($H/W=1/10$), corresponding to 
capillary number is $Ca=1.4\times 10^{-2}$. In this 
simulation $\Delta x=7 \mu m$, the viscosity of the two components $\nu=0.0167$ 
and the surface tension of the model is $\sigma=0.0244$.
After the break up, the droplet diameter is $D=350 \mu m$, so that $D/H=5$, 
in good agreement with the 
experimental findings on the volcano device 
($W/H=8$, $D/H=5.8$, $Ca=10^{-2}$) \cite{stolovicki2018throughput}.
}
\end{figure}

Here, we take as a reference the volcano device described 
in Ref. \cite{stolovicki2018throughput}
which is made of polydimethylsiloxane and used 
for producing water in oil emulsions.  
The water flows through the device inlet, and splits into hundreds 
of step-emulsifier nozzles with rectangular cross section.  
The device is submerged in a quiescent oil reservoir, each nozzles producing
a stream of micron-sized droplets.

Here, we report the preliminary simulations of a single-nozzle device,
in absence of gravity ($Bo=0$, being $Bo=\Delta \rho g D^2/\sigma$ the Bond number, 
namely the ratio between gravitational forces and surface tension.). 
As per the boundary conditions, at the inlet and outlet, we employed the 
bounce-back rule with the  momentum correction and the zero gradient boundary 
condition, respectively. Periodic boundary conditions have been 
applied along the crossflow directions.
Upon matching the governing dimensionless groups (capillary and Weber number) 
and the characteristic geometrical 
ratio $H/W$ (see the caption of fig. \ref{sequencevolc} 
for the values of the physical parameters), we are able 
to simulate the droplet break up.
After the break up, the droplet diameter is $D\sim 350\mu m$, 
corresponding to  $D/H \sim 5$, in good agreement with the 
experimental findings on the volcano 
device (\cite{stolovicki2018throughput}) ($W/H=8$, $D/H=5.8$, $Ca=10^{-2}$)).

\section{Exascale LB computing for microfluidic devices}

We have already discussed the necessity of coarse-graining to cope with the large 
separation of scales presented by microfluidic SFCs, typically from nanometers to millimetres.

In this final section, we wish to perform a "Gedanken experiment" 
and speculate on what would be possible 
if Exascale computing were with us at the time of this writing.

To begin with, let us remind that the performance of a plain LB code 
(Navier-Stokes hydrodynamics only) is of 
the order of a few MLUPS, say five, per core. 
This means that the LB code updates five million lattice sites in a CPU second.
Since a standard LB update requires about 200 Flops/site/step, 
$5$ MLUPS are moreless equivalent to $1$ Gflops/s.

Assuming a fifty percent parallel efficiency, on a Petaflop computer, we obtain 
half a million (Gigaflop to PetaFlop) speedup, yielding of the order of 2.5 TLUPS, 
which we round to i.e. 1 TLUPS, i.e. $10^{12}$ lattice updates per second.
Such performances have indeed been obtained by a few precious extreme LB simulations 
\cite{bernaschi2009muphy,bernaschi2010flexible,feichtinger2011walberla}.

With a big pinch of optimism, let us further assume that
the same fifty percent efficiency can be carried over to the Exascale, 
leading to another three orders
of magnitude speedup, i.e one $PLUPS$ rate, i.e. $10^{15}$ lattice updates per second.

Spanning five decades in space, say from $10$ nm to $1$ mm, and six in time, i.e run 
one million time steps (ten recirculation times at a typical net speed of $0.1$ lattice units), gives $10^{15} \times 10^{6} = 10 ^{21}$ lattice updates, 
to complete the full simulation of the microfluidic device at $10$ nanometer resolution.
At $1$ PLUPS, this takes $10^6$ seconds, roughly two weeks.

For a typical speed $U=1$ cm/s ($U=0.1$ in lattice units), a lattice 
spacing $\Delta x = 10$ nm corresponds to
a time-step $\Delta t = 100$ ns, so that the above simulation would span about 
$0.1$ seconds, a pretty significant time-window of the device operation.
Thus, on the prospective Exascale computer, one could simulate fractions of a second
of real-time operation of the full microfluidic device at near-nanometric resolution, in 
about two weeks elapsed time, without any coarse-graining.

Despite its extremely naive nature, this back-of-the envelope calculation 
conveys an idea of what one could possibly achieve in modeling SFC-based microfluidic 
devices, once an Exascale LB computing is with us.
Two weeks for the direct coupling between atomistic and device scale looks pretty 
close to a break-even for the landmark concept of 
"Microfluidic design at near-molecular resolution"       

Obviously, many problems must be solved along the way to make 
this dream come true, particularly
in connection with efficient parallel implementations on millions 
of parallel cores, as needed for exascale
computing.

Yet, once such problems are solved, the dream will come true, 
possibly routine operation in  decade or two. 

\section{Summary}

Summarising, we have discussed a few general features of the mesoscale modelling of
soft flowing crystals, a comparatively new state of matter with major relevance to
a number of microfluidic applications for the design of soft mesoscale materials. 

A novel variant of the Lattice Boltzmann method for multiphase flows, based on the regularisation 
of the colour-gradient scheme, augmented with a color-swap algorithm to mimic the effect of 
intermolecular repulsion, is presented. 

The new scheme has been applied to the simulation of droplet production in flow-focussing
and step-emulsifying micro devices, finding satisfactory agreement with the existing literature, both in terms
of predicting the transition from dripping-jetting-tubing regimes, and also with respect to
spacetime patterns of the droplet configurations.

Finally, we have presented a few speculations on the prospects of Exascale LB computing for
the engineering design of soft *mesoscale materials at near-molecular resolution. 

\section*{Acknowledgments}

Andrea Scagliarini and Lin Lin Fei are gratefully acknowledged for Fig. \ref{Fig:droplet_train}.
The research leading to these results has received
funding from the European Research Council under the European
Union's Horizon 2020 Framework Programme (No. FP/2014-
2020)/ERC Grant Agreement No. 739964  (\textquotedbl{}COPMAT\textquotedbl{}).


\begin{thebibliography}{10}
\expandafter\ifx\csname url\endcsname\relax
  \def\url#1{\texttt{#1}}\fi
\expandafter\ifx\csname urlprefix\endcsname\relax\def\urlprefix{URL }\fi
\expandafter\ifx\csname href\endcsname\relax
  \def\href#1#2{#2} \def\path#1{#1}\fi

\bibitem{costantini2015microfluidic}
M.~Costantini, C.~Colosi, J.~Jaroszewicz, A.~Tosato, W.~Swieszkowski,
  M.~Dentini, P.~Garstecki, A.~Barbetta, Microfluidic foaming: A powerful tool
  for tailoring the morphological and permeability properties of sponge-like
  biopolymeric scaffolds, ACS applied materials \& interfaces 7~(42) (2015)
  23660--23671.

\bibitem{shum2011multicompartment}
H.~C. Shum, Y.-j. Zhao, S.-H. Kim, D.~A. Weitz, Multicompartment polymersomes
  from double emulsions, Angewandte Chemie 123~(7) (2011) 1686--1689.

\bibitem{hamley2009ordering}
I.~Hamley, Ordering in thin films of block copolymers: Fundamentals to
  potential applications, Progress in Polymer Science 34~(11) (2009)
  1161--1210.

\bibitem{hu2014directed}
H.~Hu, M.~Gopinadhan, C.~O. Osuji, Directed self-assembly of block copolymers:
  a tutorial review of strategies for enabling nanotechnology with soft matter,
  Soft matter 10~(22) (2014) 3867--3889.

\bibitem{chremos2010ultra}
A.~Chremos, K.~Margaritis, A.~Z. Panagiotopoulos, Ultra thin films of diblock
  copolymers under shear, Soft Matter 6~(15) (2010) 3588--3595.

\bibitem{cheng2004nanostructure}
J.~Y. Cheng, A.~M. Mayes, C.~A. Ross, Nanostructure engineering by templated
  self-assembly of block copolymers, Nature materials 3~(11) (2004) 823.

\bibitem{garstecki2006flowing}
P.~Garstecki, G.~M. Whitesides, Flowing crystals: nonequilibrium structure of
  foam, Physical review letters 97~(2) (2006) 024503.

\bibitem{succi2018lattice}
S.~Succi, The Lattice Boltzmann Equation: For Complex States of Flowing Matter,
  Oxford University Press, 2018.

\bibitem{kruger2017lattice}
T.~Kr{\"u}ger, H.~Kusumaatmaja, A.~Kuzmin, O.~Shardt, G.~Silva, E.~M. Viggen,
  The lattice boltzmann method, Springer International Publishing 10 (2017)
  978--3.

\bibitem{gunstensen1991lattice}
A.~K. Gunstensen, D.~H. Rothman, S.~Zaleski, G.~Zanetti, Lattice boltzmann
  model of immiscible fluids, Physical Review A 43~(8) (1991) 4320.

\bibitem{leclaire2012numerical}
S.~Leclaire, M.~Reggio, J.-Y. Tr{\'e}panier, Numerical evaluation of two
  recoloring operators for an immiscible two-phase flow lattice boltzmann
  model, Applied Mathematical Modelling 36~(5) (2012) 2237--2252.

\bibitem{brackbill1992continuum}
J.~Brackbill, D.~B. Kothe, C.~Zemach, A continuum method for modeling surface
  tension, Journal of computational physics 100~(2) (1992) 335--354.

\bibitem{landau1959course}
L.~Landau, E.~Lifshitz, Course of theoretical physics. vol. 6: Fluid mechanics,
  London, 1959.

\bibitem{liu2012three}
H.~Liu, A.~J. Valocchi, Q.~Kang, Three-dimensional lattice boltzmann model for
  immiscible two-phase flow simulations, Physical Review E 85~(4) (2012)
  046309.

\bibitem{reis2007lattice}
T.~Reis, T.~Phillips, Lattice boltzmann model for simulating immiscible
  two-phase flows, Journal of Physics A: Mathematical and Theoretical 40~(14)
  (2007) 4033.

\bibitem{latva2005diffusion}
M.~Latva-Kokko, D.~H. Rothman, Diffusion properties of gradient-based lattice
  boltzmann models of immiscible fluids, Physical Review E 71~(5) (2005)
  056702.

\bibitem{d1995two}
U.~d'Ortona, D.~Salin, M.~Cieplak, R.~B. Rybka, J.~R. Banavar, Two-color
  nonlinear boltzmann cellular automata: Surface tension and wetting, Physical
  Review E 51~(4) (1995) 3718.

\bibitem{montessori2018regularized}
A.~Montessori, M.~Lauricella, M.~La~Rocca, S.~Succi, E.~Stolovicki, R.~Ziblat,
  D.~Weitz, Regularized lattice boltzmann multicomponent models for low
  capillary and reynolds microfluidics flows, Computers \& Fluids 167 (2018)
  33--39.

\bibitem{dupin2006simulation}
M.~M. Dupin, I.~Halliday, C.~M. Care, Simulation of a microfluidic
  flow-focusing device, Physical review E 73~(5) (2006) 055701.

\bibitem{montessori2015lattice}
A.~Montessori, P.~Prestininzi, M.~La~Rocca, S.~Succi, Lattice boltzmann
  approach for complex nonequilibrium flows, Physical Review E 92~(4) (2015)
  043308.

\bibitem{zhang2006efficient}
R.~Zhang, X.~Shan, H.~Chen, Efficient kinetic method for fluid simulation
  beyond the navier-stokes equation, Physical Review E 74~(4) (2006) 046703.

\bibitem{stolovicki2018throughput}
E.~Stolovicki, R.~Ziblat, D.~A. Weitz, Throughput enhancement of parallel step
  emulsifier devices by shear-free and efficient nozzle clearance, Lab on a
  Chip 18~(1) (2018) 132--138.

\bibitem{cubaud2008capillary}
T.~Cubaud, T.~G. Mason, Capillary threads and viscous droplets in square
  microchannels, Physics of Fluids 20~(5) (2008) 053302.

\bibitem{garstecki2005mechanism}
P.~Garstecki, H.~A. Stone, G.~M. Whitesides, Mechanism for flow-rate controlled
  breakup in confined geometries: A route to monodisperse emulsions, Physical
  review letters 94~(16) (2005) 164501.

\bibitem{costantini2014highly}
M.~Costantini, C.~Colosi, J.~Guzowski, A.~Barbetta, J.~Jaroszewicz,
  W.~{\'S}wi{\k{e}}szkowski, M.~Dentini, P.~Garstecki, Highly ordered and
  tunable polyhipes by using microfluidics, Journal of Materials Chemistry B
  2~(16) (2014) 2290--2300.

\bibitem{bernaschi2009muphy}
M.~Bernaschi, S.~Melchionna, S.~Succi, M.~Fyta, E.~Kaxiras, J.~K. Sircar,
  Muphy: A parallel multi physics/scale code for high performance bio-fluidic
  simulations, Computer Physics Communications 180~(9) (2009) 1495--1502.

\bibitem{bernaschi2010flexible}
M.~Bernaschi, M.~Fatica, S.~Melchionna, S.~Succi, E.~Kaxiras, A flexible
  high-performance lattice boltzmann gpu code for the simulations of fluid
  flows in complex geometries, Concurrency and Computation: Practice and
  Experience 22~(1) (2010) 1--14.

\bibitem{feichtinger2011walberla}
C.~Feichtinger, S.~Donath, H.~K{\"o}stler, J.~G{\"o}tz, U.~R{\"u}de, Walberla:
  Hpc software design for computational engineering simulations, Journal of
  Computational Science 2~(2) (2011) 105--112.

\end{thebibliography}

\end{document}